# CW-pumped telecom band polarization entangled photon pair generation in a Sagnac interferometer

Yan Li,[1,2] Zhi-Yuan Zhou,[1,2] Dong-Sheng Ding,[1,2] and Bao-Sen Shi[1,2,3]

[1]*Key Laboratory of Quantum Information, University of Science and Technology of China, Hefei, Anhui 230026, China*

[2]*Synergetic Innovation Center of Quantum Information & Quantum Physics, University of Science and Technology of China, Hefei, Anhui 230026, Chna*

[3]*zyzhouphy@mail.ustc.edu.cn*

[4]*drshi@ustc.edu.cn*

A polarization entangled photon pair source is widely used in many quantum information processing applications such as teleportation, quantum swapping, quantum computation and high precision quantum metrology. Here, we report on the generation of a continuous-wave pumped degenerated 1550 nm polarization entangled photon pair source at telecom wavelength using a type-II phase-matched periodically poled $KTiOPO_4$ crystal in a Sagnac interferometer. Hong-Ou-Mandel-type interference measurement shows the photon bandwidth of 2.4 nm. High quality of entanglement is verified by various kinds of measurements, for example two-photon interference fringes, Bell inequality and quantum states tomography. The wavelength of photons can be tuned over a broad range by changing the temperature of crystal or pump power without losing the quality of entanglement. This source will be useful for building up long-distance quantum networks.

**1. Introduction**

Entangled photon sources are basic platforms for quantum optical experiments and quantum information processing tasks like quantum key distribution [1-3], quantum teleportation [4-6], photonic frequency conversion [7], and quantum computations [8]. A polarization entangled photon source is one of the most important entangled photon sources that have been studied for decades of years. To date, the most successful method in generating polarization entangled photon pairs is based on the spontaneous parametric down-conversion process (SPDC) in nonlinear crystals. In the literatures [9-18], people generate polarization entangled photons using different crystals with different experimental configurations. In the early times, polarization entangled photons are created using birefringence phase matching (BPM) crystals, a type-II phase matched BBO crystal is used to create a polarization entangled photon source in the first practical and effective experiment, in which orthogonal polarization entangled photons are emitted at the intersection cones [9]. The significant progress in nonlinear crystal fabrication makes a quasi-phase matching (QPM) crystal a better choice for researcher in many nonlinear optics applications. The most important merits of using QPM crystals in generation photon pairs is its high spectral brightness in contrast to BPM crystals, due to its large effective nonlinear coefficient and long allowable interaction length.

Recently, to generate polarization entangled photon pairs by placing a QPM crystal inside a Sagnac interferometer configuration has been demonstrated to be superior than other configurations. The merits to use the Sagnac interferometer configuration are its compactness, high stability and high brightness. Kim *et al*, prepared the first polarization entangled source by pumping a periodically poled KTiOPO4 (PPKTP) with a continuous-wave (CW) laser at 405nm in a Sagnac-loop in 2006 [19, 20] following the idea of Ref. 10. Then, a pulsed polarization entangled source at 780 nm based on this configuration was developed by Kuzucu and Wong in 2008 [21]. Now, polarization entangled sources based on QPM crystals in a Sagnac configuration have become a basic tool for many experiments [22-28]. In the early experiments, the wavelengths of the photons generated are in visible range, therefore these photons are not suitable for long distance quantum communications in fiber. Only recently, telecom band polarization entangled photon sources are developed. A pulsed polarization entangled source at 1584 nm based on type-II PPKTP was demonstrated by Jin *et al*, in 2014 [18]. However, the CW pump polarization entangled source at telecom band based on a single type-II PPKTP in a Sagnac-loop configuration has not been demonstrated yet.

Spectral indistinguishibility between the signal and idler photon is of vital importance in multi-entangled-source based experiments [29-31]. For a CW pumped degenerate photon pair source, the spectral of signal and idler photon can be tuned to overlap nearly perfectly at any pump wavelength, and high visibility for the two-photon Hong-Ou-Mandel (HOM) interference can be obtained without strict spectral filtering. But for a pulsed pump photon pair source, high purity of the photon source is usually obtained near the group velocity matched (GVM) wavelength. In [32], a high purity photon pair source at wavelength ranging from 1460 nm to 1675 nm is obtained near the GVM wavelength of 1584 nm. Another difference between the CW pumped and the pulse pumped photon pair sources are the multi-photon generation probability at the same pump power, the peak power for a pulse is very high, so the multi-photon generation probability of a pulse pumped scheme is much greater than that in a CW pumped scheme.

Besides high spectral indistinguishability of a CW pumped photon pair source, another important merit of our CW pumped Sagnac interferometer-based source is that such a high-quality polarization entangled source at telecom wavelength is suitable for long distance transmission in low loss fiber, which will be very important for fiber based quantum communication systems. In addition, the main aim to build this source is to convert the entanglement from photons' polarization degree of freedoms to orbital angular momentum (OAM) degree of freedoms [28], which will be used for quantum frequency conversion in our future experiments [33].

In this work, we prepare a CW-pumped 1550 nm telecom wavelength polarization entangled photon pair source with a single type-II PPKTP crystal in a Sagnac-loop configuration. Various measurements are performed to characterize the quality of the entangled source. The two-photon HOM interference with 95.3%±1.6% visibility yields 2.4 nm photon bandwidth. High HOM interference visibility keeps unchanged even the central wavelength of the photon is tuned over 20 nm. The two-photon Bell-type interference fringe at $45^0$ basis has visibility of 96.4%±2.0%. The Bell-type visibilities keep unchanged even the pump power is varied from 15 mW to 120 mW or the temperature of the crystal is changed from 15℃ to 55℃. The measured S parameter of Clauser-Horne-Shimony-Holt (CHSH) inequality is 2.63± 0.08, which violates the inequality with 8 standard deviations. We also perform state tomography of the entangled state, obtain state fidelity of 0.935± 0.021. These results clearly show the high performance of our entangled source.

## 2. Experimental setups

The experimental setup is showed in Fig. 1, where (a) shows the experimental setup for photon pair preparation and (b) for two-photon HOM interference experiments. The CW pump laser at 775 nm is from a Ti: sapphire laser (Coherent MBR 110), the pump beam is collected to single mode fiber before entering the Sagnac-loop. A quarter wave plate (QWP) and a half wave plate (HWP) are used to control the phase and intensity of the pump beams in the Sagnac-loop. The pump laser is focused by a lens with focus length of 200 mm, whose beam waist is about 40 μm at the center of the PPKTP crystal. The type-II PPKTP (Raicol crystals) crystal has size of 1 mm×2 mm×10 mm, with a periodical poling period of 46.2 μm. The temperature of the PPKTP crystal is controlled by a homemade temperature controller with a stability of 2mK. The horizontal and vertical parts of the pump beam are separated by a double polarization beam splitter (DPBS). The vertical polarized part of the pump beam is rotated to horizontal polarization by a double half wave plate (DHWP) before it enters the PPKTP crystal. The counter-propagating orthogonal polarized photon pairs generated are recombined at the DPBS and collected into a single mode fiber by using a lens set consisting of two lenses with different focus length of 100 mm and 50 mm respectively at each output port of the DPBS. The pump beam is removed with long pass filter (FELH1400). We use one HWP and one polarizer (P) to perform correlation measurement. For quantum states tomography, QWP, HWP and P are used at each output port. The collected photons are detected by InGaAs single photon avalanche detectors (APD1 and APD2).

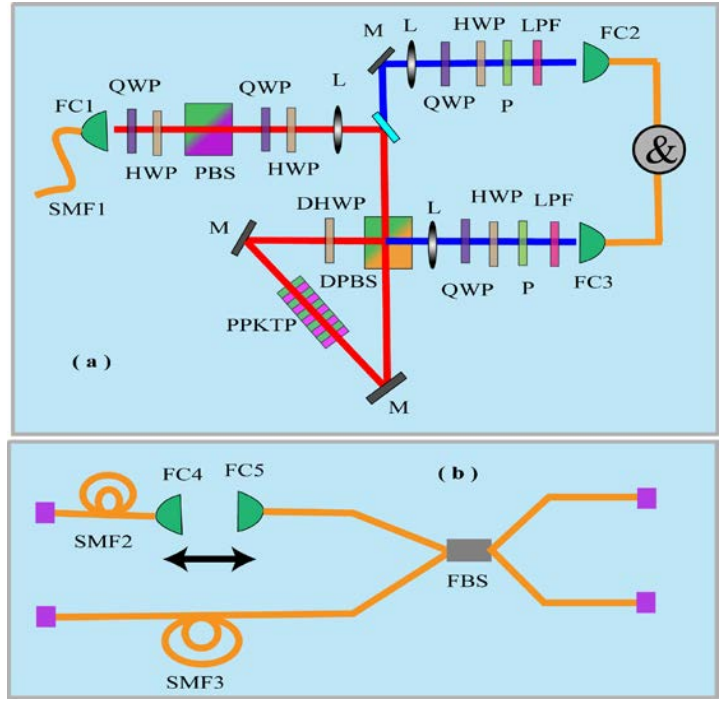

Fig. 1. (a) Experimental setup for the polarization entangled source in Sagnac-loop configuration; (b) experimental setup for two-photon HOM interference. L: lens; Q(H)WP: quarter (half) wave plate; M:mirror; P: polarizer; DM: dichromatic mirror; LPF: long pass filter; FC1-5: fiber coupler; DHWP: double half wave plate for the pump at 780nm and SPDC photons at 1560nm; DPBS: double polarization beam splitter for the pump and SPDC photons at 780nm and 1560nm respectively; PBS: polarization beam splitter; PPKTP: periodically poled KTP crystal; SMF1-3:single mode fibers; FBS: fiber beam splitter.

## 3. Experimental results

Before charactering the entanglement properties of the photon pair source, we perform two-photon HOM interference firstly to measure the bandwidth and degenerate temperature of the signal and idler photon. In this measurement, the pump beam's polarization is controlled to be vertical, and pump only circulates counter-clock wise in the Sagnac-loop. The polarizer at each port is removed, signal and idler photons collected by fiber couplers FC2 and FC3 are sent to the setup depicted in Fig. 1(b). Two photons are combined using a fiber beam splitter (FBS); in one arm of the interferometer, we create an air gap using fiber couplers FC4 and FC5, the length of air gap can be changed by a one dimensional translator with a step of 10 μm. The balance position of the interferometer is obtained by interfering a 50 ps attenuated pulsed laser at 1550 nm. The two outputs of the interferometer are connected to single photon detectors (APD1 and ADP2. APD1 is from Lightwave Princeton, with 90 MHz trigger rate, 1ns detection window and 15% detection efficiency, the dark cont probability is about $5.6\times10^{-6}$ per gate. APD2 is from Qask, 8% detection efficiency, 2.5 ns detection window and the dark count probability is about $2.0\times10^{-5}$ per gate.) Photons are delayed with a 200-m long single mode fiber (SMF) before they are detected with APD2, the output of APD1 is used to trigger the detector APD2 for coincidence measurements. A digital delay generator is used to compensate the time difference between the two photons.

The experimental result for the HOM interference curve is showed in Fig. 2. In this measurement, the pump beam power is 50 mW, the single count is about 9 kcps, and the dark count is 500 cps. The visibility for the interference is defined as $V=(C_{max}-C_{min})/C_{max}$, the measured raw visibility is $95.3\%\pm1.6\%$, the full wave half maximum (FWHM) width of the fitted curve gives two photon coherent length $l_c$ of 0.44 mm. The bandwidth of the photons is related to the coherent length with the formula of $\Delta\lambda=1.39\lambda^2/\pi l_c$, where $\lambda$ is the central wavelength (1550 nm) of the photon pair. Therefore the measured bandwidth of the signal and idler photon is $2.4\pm0.1$ nm, which is agreement with the theoretical calculation value of 2.4 nm with the dispersion function gives in [34].

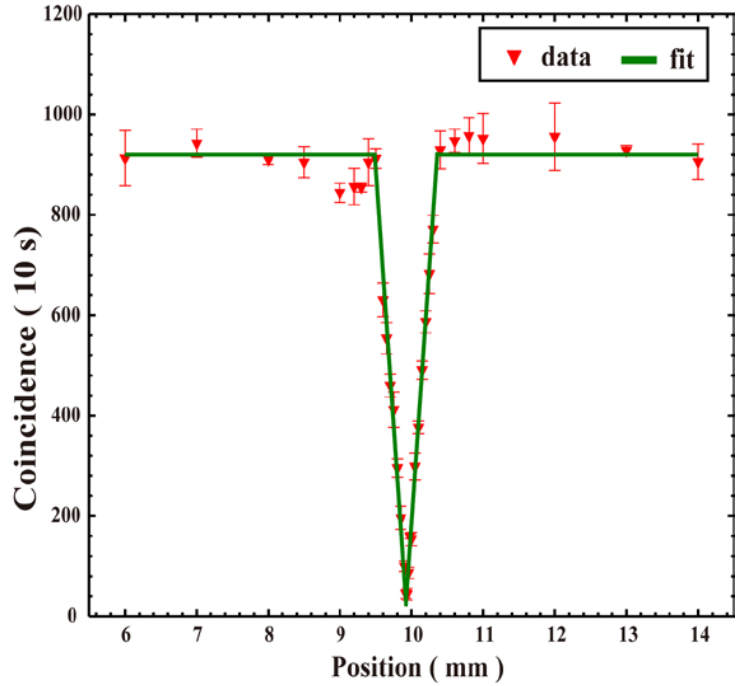

Fig. 2. Coincidence counts in 10 second as a function of the displacement of the adjustable air gap. Error bars are estimated from multiple measurements; the experimental data is fitted using triangle function.

We also measure the HOM interference at other central wavelength of photon by tuning the pump wavelength and the crystal temperature, the results are showed in table 1. Table 1 show that the HOM visibility keeps nearly constant over wavelength ranging from 1540 nm to 1560 nm. Therefore the signal and idler spectra overlap perfectly at a broad wavelength range.

Table 1. HOM interference measurements at different signal wavelengths.

| Signal photon Wavelength (nm) | Temperature (□) | Single count rate (kcps) | Visibility (%) |
|---|---|---|---|
| 1539.90 | 51.0 | 8.0 | 96.6±1.3 |
| 1545.47 | 41.0 | 8.5 | 97.5±0.5 |
| 1550.09 | 32.0 | 9.0 | 95.3±1.6 |
| 1554.72 | 27.0 | 9.4 | 96.7±0.3 |
| 1560.31 | 17.5 | 8.5 | 95.0±0.8 |

After performing HOM interference for the photon pair, now we will characterize the entanglement photon pair source. The output state of the Sagnac interferometer can be expressed as

$$|\Phi\rangle = \frac{1}{\sqrt{2}}(|HV\rangle + e^{i\theta}|VH\rangle), \qquad (1)$$

To obtain Eq. (1), we should balance the photon generate rate between two circulation directions. The relative phase $\theta$ can be determined by the positions of the QWP, HWP at the input port of the interferometer and the position of the crystal inside the Sagnac loop by changing the Gouy phase [35]. By rotating QWP and HWP in the pump beam, we can obtain the state

$$|\Phi^+\rangle = \frac{1}{\sqrt{2}}(|HV\rangle + |VH\rangle). \qquad (2)$$

The pump power is fixed to be 60 mW and the pump beam wavelength is tuned to be 775.04 nm, the temperature of the crystal is kept at 32℃. Then we perform polarization correlation measurement between the two photons. We fix the angle of the HWP at one of the output at 0 and 22.5 degree respectively, and measure the two-photon coincidence as a function of the HWP rotation angle at the other output port. The experimental results are showed in Fig. 3, the raw visibilities are 96.7%±1.4% and 96.4%±2.0% respectively. The entanglement visibility is defined as $V = (C_{max} - C_{min})/(C_{max} + C_{min})$, where $C_{max}$ and $C_{min}$ are the maximum and minimum coincidence count respectively. The visibility is calculated from the sinusoidal function fits of the experimental data sets. The time for each coincidence measurement is 10 seconds, the single count rate at each port is about 7.5 kcps, the dark count rate is about 500 cps.

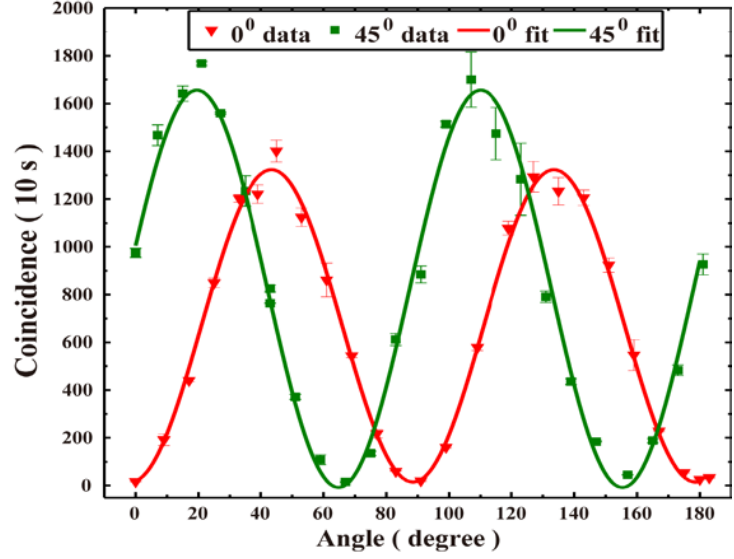

Fig. 3. Two photon coincidence in 10s as a function of the angle the two HWPs. The background dark coincidence is not subtracted; error bars are obtained from multiple measurements; the data sets are fitted using sinusoidal function.

The two-photon interference visibilities obtained experimentally with greater than 71% clearly indicate the violation of Bell inequality. To further characterize the quality of the present polarization entangled source, we also calculate the Bell S parameter for different settings of the polarizers [36]. The measured S parameter in 10 s for state $|\Phi^+\rangle$ is $2.63\pm 0.08$, which violates the Bell inequality with 8 standard deviations.

The average filtering and coupling efficiencies $\alpha_1, \alpha_2$ are 0.82 and 0.32 for each photon respectively. We assume the collection efficiencies for signal and idler photons are the same, the collection efficiency for signal photon and idler photon are both $\alpha_1\alpha_2$, and the overall photon pair collection efficiency is $(\alpha_1\alpha_2)^2$. The duty cycle $d$ of the single photon detector is 0.09. Accounting for all the losses including collection losses and detection losses, the inferred spectral brightness $B_{Inferred} = 2N_c/(\alpha_1^2\alpha_2^2 d\eta_1\eta_2 P\Delta\lambda)$ is about $3.0\times 10^4 (\text{s}\cdot \text{mW}\cdot \text{nm})^{-1}$, where $N_c = 150$ cps is the coincidence rate per second, $\eta_1 = 0.15$ and $\eta_2 = 0.08$ are the detection efficiencies of APD1 and APD2 respectively, $P = 60$ mW is the pump power and $\Delta\lambda = 2.4$ nm is the bandwidth of the signal and idler photons. We also can calculate the experimentally detected spectral brightness with the formula $B_{Detected} = 2N_c / (P\Delta\lambda)$, which gives a detected spectral brightness of 2 $(s \cdot mW \cdot nm)^{-1}$, the detected spectral brightness can be greatly enhanced by using superconducting nanowire single photon detectors.

To precisely know which Bell state is generated, we also carry out quantum tomography. The experimental density matrix $\rho_{exp}$ reconstructed using maximum-likihood estimation method [37] is showed in Fig. 4, the fidelity of the reconstructed density matrix to the ideal Bell state $|\Phi^+\rangle$ is defined as $F = \langle\Phi^+|\rho_{exp}|\Phi^+\rangle$, we estimate the fidelity of our present source is $0.935 \pm 0.021$. To obtain this high fidelity, the angles of wave plates at each port are calibrate very carefully by using a CW coherent laser beam at 1550 nm. In addition, the pump laser is reshaped by a single mode fiber, therefore has high beam quality, thus the fiber collection efficiency of the photon pairs can be improved. These values measured indicate that the source we generate is highly entangled.

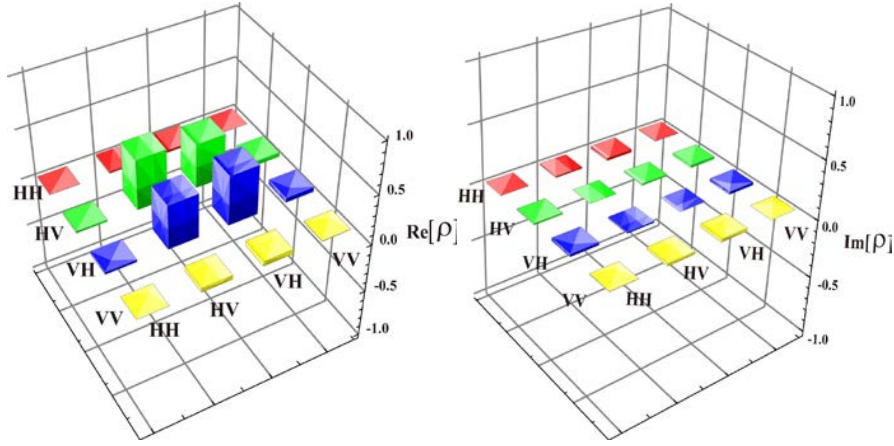

Fig. 4. Real (left) and imaginary (right) parts of the experimental reconstructed density matrix for the polarized entangled source via maximum likihood estimation method.

In addition to these measurements, we also measure the visibilities for two-photon interference at $45^0$ basis by using different pump power or by setting different crystal temperature. The results are showed in Fig. 5. Fig. 5(a) shows the visibilities as a function of pump beam power, the crystal temperature kept unchanged for the measurements. Fig. 5(b) shows the visibilities at different crystal temperatures, the pump power is kept to be 60 mW. We conclude from Figs. 5(a) and 5(b) that the visibilities keep at a very high level no matter temperature and pump beam power. Therefore our source is robust against various experimental parameters.

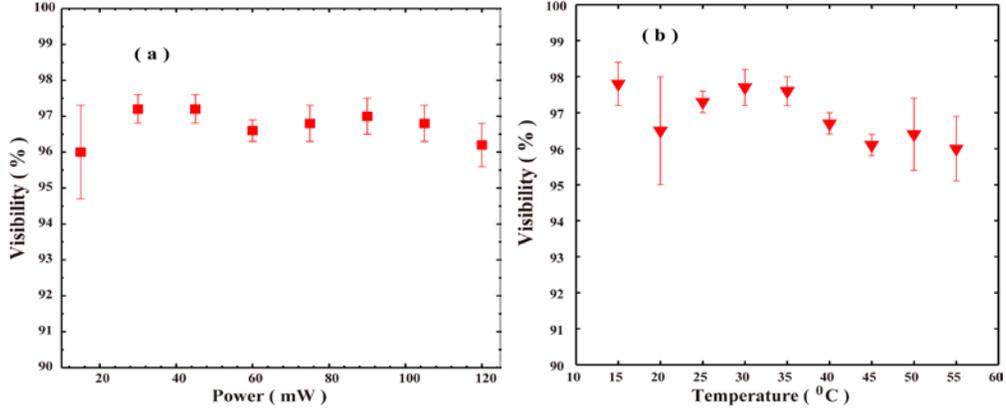

Fig. 5. (a)Visibilities as a function of pump beam power, the power is varying from 15 mW to 120 mW; (b) visibilities as a function of crystal temperature, the temperature is tuning from 15 ℃ to 55 ℃. The error bars are obtained from multiple measurements.

## 4. Outlook and discussion

By comparing our source with previous demonstration using a pulse pumped Sagnac-loop configuration at telecom band [18], the total photon collection efficiency is enhanced by using high transmission long pass filters and carefully aligning the Sagnac-loop. The parameters characterizing the quality of the source is closed to that in [18] even with poor single photon detectors. The performance of the source will be much better if high efficient superconducting nanowire single photon detectors are used, as the coincidence rate will be much higher and system statistic error will be much lower. We also find that the raw visibility keeps nearly unchanged against pump power, while the raw visibility decreases very fast in ref. [18] in the same power range. In addition, we also demonstrate the temperature tuning ability of the source, which is not shown in [18].

Since the PPKTP crystal working at the demonstrated wavelength regime has broad tuning ability [16], so the source central wavelength can be widely-tuned by using broad band DPBS and DHWP in Fig. 1(a). The polarization entangled source we prepare will be used in many quantum information and quantum communication experiments in our lab in the future. For example, we will convert polarization entangled source to two dimensional OAM entangled source based on the method reported in [28], the telecom band OAM entangled source will be used for quantum frequency up-conversion experiments. In addition, such source is also suitable for entangled-based quantum key distribution experiments in free space or optical fibers.

## 5. Conclusion

For summary, a high-performed CW polarization entangled photon source at telecom band is created. 95.3%±1.6% visibility in HOM interference shows the photon bandwidth of 2.4 nm. Raw visibility 96.4%±2.0% in Bell-type interference is obtained. The reconstructed density matrix of the state has fidelity of 0.935±0.021. The visibilities keep unchanged against broad tuning range of pump power and crystal temperature. These evidences show that the present source will be very promising in quantum information and communication experiments in our lab in the future.

**Acknowledgments**

This work was supported by the National Fundamental Research Program of China (Grant No. 2011CBA00200), the National Natural Science Foundation of China (Grant Nos. 11174271, 61275115, 10874171).